  \documentclass[12pt,a4paper]{article}
  \usepackage{amsfonts}
  \newcommand{\field}[1]{\mathbb{#1}}
  \newcommand{\rbb}{\field{R}}
  \newcommand{\cbb}{\field{C}}
  \newcommand{\wed}{\!\wedge\!}         
  \newcommand{\tr}{\mathrm{Tr}\,}       
  \newcommand{\nc}{\kappa}              
  
  \newcommand{\beq}{\begin{equation}}
  \newcommand{\eeq}{\end{equation}}

  \newcommand{\fb}{\mathbf{E}}          
  \newcommand{\fp}{\mathbf{p}}          
\begin{document}
\title{A Remark on Topological Charges \\over the Fuzzy Sphere}
\author{
Harald Grosse\footnote{E-mail: grosse@thp.univie.ac.at} $\:$and
Christian W. Rupp\footnote{E-mail: cwrupp@itp.uni-leipzig.de}}
\date{\small
Institut f\"ur Theoretische Physik, Universit\"at Wien, \\
Boltzmanng. 5, A--1090 Wien, Austria, \\
Institut f\"ur Theoretische Physik, Universit\"at Leipzig \\
Augustusplatz 10--11, D--04109 Leipzig, Germany}
\maketitle
\begin{abstract}
\noindent
We determine the Chern characters of two projective modules over the fuzzy
sphere and calculate the corresponding topological charges (Chern numbers).
These turn out to have corrections---compared to the commutative limit---induced
by the noncommutative structure of the three coordinates.
\end{abstract}
\section{Introduction and Overview}
Classical gauge field theories exhibit interesting features connected with the
geometry and topology of nontrivial fiber bundles (over space or space--time).
Examples are mono\-pole and in\-stan\-ton solutions. In noncommutative geometry
the situation is similar since the algebraic aspect of nontriviality is nothing 
but the projectivity of certain algebra modules.

For the fuzzy sphere this 
case has first been analyzed by one of us (HG) et. al. in \cite{grosse96a}
(see also \cite{grosse98}),  leading to scalar and spinor field configurations 
in monopole backgrounds. A different approach using spectral triples and their 
Dirac operators has been used in \cite{balachandran00} and  
\cite{balachandran01}. Recently \cite{valtancoli} there have been attempts to 
calculate Chern numbers applying a noncommutative version of the projectors for 
the 2--sphere used in \cite{landi00} and \cite{landi01}. In our present analysis 
we use this setup to show that the Chern characters obtained in this way
give rise to non-integer (fuzzy) topological charges, becoming integer in the 
commutative limit.

In section \ref{chern} we briefly review the Chern character on 
projective modules over algebras equipped with a differential calculus, 
primarily to settle notation. Further information can be found e.g. in 
\cite{landi97}, \cite{karoubi} and \cite{connes94}. Then in section \ref{s2} we 
review the complex line bundles over the 2--sphere,
and in section \ref{fuzzyline} we generalize the bundles with topological charge
$\pm 1$ to the fuzzy sphere, resulting in fuzzy line bundles, 
their Chern characters and Chern numbers.
\section{The Chern Character} \label{chern} 
Let $\mathcal A$ be a complex unital not necessarily commutative $C^{\star}$-algebra 
and denote ${\mathcal A}\otimes {\cbb}^{n}$ by ${\mathcal A}^{n}$. Then any 
projector (selfadjoint idempotent) $p\in M_n(\mathcal A)$, the $n\!\times\! 
n$-matrices with coefficients in $\mathcal{A}$, defines a (finitely generated) 
projective right $\mathcal A$-module $E = p{\mathcal A}^{n}$. Elements $\psi$ of 
$E$ can be viewed as $\psi\in {\mathcal A}^{n}$ with $p\psi=\psi$. If $\mathcal 
A$ is further endowed with a differential calculus $\left( {\Omega^*({\mathcal 
A}), d}\right)$, the Grassmann connection $\nabla: E\rightarrow
E\otimes_{\mathcal A}\Omega^1({\mathcal A})$ of $E$ is defined by
$\nabla = p\circ d$. It satisfies $\nabla(\psi f)=(\nabla\psi)f +
\psi\otimes_{\mathcal A}df$ for all $f \in \mathcal A$, $\psi\in E$. After
extending $\nabla$ to $E\otimes_{\mathcal A}\Omega^1({\mathcal A})$ one can 
define the $\mathcal A$-linear map
\[
\nabla^2:E\rightarrow E\otimes_{\mathcal A}\Omega^2({\mathcal A}),
\]
called the curvature of $\nabla$. Evaluating $\nabla^2$ one finds
$\nabla^2=p(dp)(dp)$.
The Chern character of $E$ is the set of 
\beq
\mathrm{Ch}_q(p):=\frac{1}{q!}\tr p(dp)^{2q},
\eeq
which are a cocycles and provide equivalence classes in $H^{2q}(\mathcal{A})$. 
$\mathrm{Ch}_0(p)=\tr p$ simply gives the rank of the module. In the case of the 
2--sphere and the fuzzy sphere the highest non-vanishing component of the Chern
character is $\mathrm{Ch}_1$. If $\mathcal{A}$ is a commutative algebra and
$E_1$ and $E_2$ are both projective $\mathcal{A}$--modules, then the first 
component of the Chern character on the module tensor product has the property
\beq
\mathrm{Ch}_1(E_1\otimes_{\mathcal{A}} E_2) = 
\sum_{i+j=1}\mathrm{Ch}_i(E_1)\mathrm{Ch}_j(E_2), \label{chernclass}
\eeq
as it is the case for bundles of higher topological charge over the 2--sphere.
\section{Line Bundles over the 2--Sphere} \label{s2} In this section we give an 
instant derivation of all inequivalent complex line bundles over $S^2$. Define 
the 2-sphere by
\[
(x_1,x_2,x_3)\in \rbb^3 \quad\mbox{with}\quad (x_1)^2+(x_2)^2+(x_3)^2=1
\]
and denote by ${\mathcal A}=C^{\infty}(S^2)$ the commutative $\cbb$-algebra
of smooth functions on it. The Bott projector $p=p^2=p^\dagger$ is
\beq
p=\frac{1}{2}(1+\sigma_a x_a)\in M_2({\mathcal A}),
\eeq
a $2\!\times\!2$-matrix with coefficients in $\mathcal A$, where $\sigma_a$ are 
the three Pauli matrices. $p$ defines a line bundle---note that $\tr p = 
1$---over $S^2$ through the right $\mathcal A$-module of smooth sections $E= 
p{\mathcal A}^2$. The first component $F\in \Omega^2(S^2)$ of the Chern character
of $E$ is given by
\begin{eqnarray}
F & = & \tr p(dp)(dp) = \frac{1}{8}\tr (1+\sigma_ax_a)(\sigma_bdx_b)(\sigma_cdx_c)
                   \nonumber \\
  & = & \frac{1}{8} \tr (\sigma_b \sigma_c + \sigma_a x_a \sigma_b \sigma_c)dx_b 
        \wed dx_c. \nonumber
\end{eqnarray}
Using $\sigma_b \sigma_c=\delta_{bc}+i\epsilon_{bcd}\sigma_d$ and the 
tracelessness of the Pauli matrices this expression can be seen to be equal to
\beq
F = \frac{1}{4}i\epsilon_{abc}x_adx_b\wed dx_c.
\eeq
Since $\frac{1}{2}\epsilon_{abc}x_adx_b\wed dx_c = \sin\theta d\theta\wed d\phi$ 
it holds that
\[
\int_{S^2} \frac{1}{8\pi} \epsilon_{abc}x_adx_b\wed dx_c =1,
\]
and therefore the first Chern number (topological charge) $c_1(p)$ of the bundle 
$E$ is found to be
\beq
c_1(p)= \frac{1}{2\pi i}\int_{S^2}F = 1.
\eeq
Notice that $c_1(p^{\mathrm t})=-1$. Bundles of higher topological charge can 
be constructed through the tensor product $E_{(k)}:=E\otimes_{\mathcal{A}}
\cdots\otimes_{\mathcal{A}}\! E$. In fact $E_{(k)}=p_k(\mathcal{A})^{2^k}$, 
where $p_k:=p\otimes_{\mathcal{A}}\cdots\otimes_{\mathcal{A}} p \in M_{2^k}
(\mathcal A)$ is again a 
projector with $\tr p_k=1$, i.e. the bundles are complex line bundles, 
$\mathrm{rank}_{\cbb}\,E_{(k)}=1$. Now
\[
dp_k=\sum_{i=1}^k p\otimes_{\mathcal{A}}\cdots\otimes_{\mathcal{A}}
dp\otimes_{\mathcal{A}}\cdots\otimes_{\mathcal{A}} p,
\]
where $dp$ appears once at the $i$-th position.
Since $p(dp)p=0$ and $\tr(a\otimes_{\mathcal{A}} b)=
\tr a \,\tr a$ we get for the Chern character of $E_{(k)}$
\beq
F_{k}= \tr p_k(dp_k)(dp_k)=k\tr p(dp)(dp),
\eeq
i.e. simply $k$ times the Chern character of charge one, and consequently for the 
associated topological charges
\beq
c_1(p_k)=\frac{1}{2\pi i}\int_{S^2}F_{k}=k.
\eeq
The transposed projectors provide the negative Chern numbers 
$c_1(p^{\mathrm{t}}_k)=-k$. 
\section{Fuzzy Line Bundles} \label{fuzzyline} The fuzzy sphere \cite{madore92},
\cite{madore99} 
is defined by the relations \beq [X_a,X_b]=i\nc \epsilon_{abc} X_c,\:\:\:\: 
\sum_{a=1}^3 (X_a)^2=1, \label{fuzzy} \eeq where $\{X_a\}_{a=1}^3$ generate the 
irreducible spin $j$ representation of $su(2)$ and $\nc$ is not a continuous 
parameter but restricted to \beq
\nc = \frac{1}{\sqrt{j(j+1)}}.
\eeq
The associative $\cbb$-algebra generated by the $X_a$'s is $M_N(\cbb)=:{\mathcal A}_N$,
the algebra of $N\!\times\! N$-matrices with $N=2j+1$. The differential calculus 
on $\mathcal{A}_N$ to make it the fuzzy sphere is derivation based: One chooses 
the three derivations defined by $e_a:= (1/\nc)\,\mathrm{ad}\,X_a$, satisfying
$[e_a,e_b]=i\epsilon_{abc}e_c$, as analogue of the set of vector fields on the  
2-sphere. One-forms $df$ for $f\in \mathcal{A}_N$ are given through
\[
df(u)=u(f), \:\mbox{ where } u=u^ae_a,
\]
and $u^a\in\cbb$. Particularly $dX_a(u)=u(X_a)$. One can choose a
basis $\theta^a$ of $\Omega^1(\mathcal{A}_N)$ completely determined by
$\theta^a\!(e_b)=\delta^a_b$ such that $df=e_a(f)\theta^a$.

To proceed analogously to the commutative case we define
$\fp\in M_2({\mathcal A}_N)$ through
\beq
\fp:= \alpha + \beta\,\sigma_a\otimes X_a,
\eeq
with $\alpha, \beta \in \rbb$. Then $\fp$ is a projector if and only if
\beq
\beta=\pm\frac{1}{\sqrt{4+\nc^2}}\mbox{ and } \alpha=\frac{1+\beta\nc}{2}.
\eeq
Accordingly, with this choice for $\alpha$ and $\beta$,
$\fp$ defines a  projective right $\mathcal{A}_N$-module $\fb=\fp\mathcal{A}^2_N$.
Note that
\[
\tr \fp=1+\beta\nc=1\pm\frac{1}{N},
\]
which is unequal $1$ as long as $\nc\neq 0$.
We shall refer therefore to this noncommutative ``line'' bundles as fuzzy line
bundles, their ``fiber dimension'' deviating slightly\footnote{We prefer to 
think of the noncommutativity as $\nc\ll 1$.} from $1$. The first component
$\mathbf{F}$ of the Chern character of $\fb$ is given by
\[
\mathbf{F}= \tr \fp(d\fp)(d\fp) = \tr \beta^2(\alpha +
\beta\,\sigma_a\otimes X_a)(\sigma_b\otimes dX_b)(\sigma_c\otimes dX_c).
\]
A calculation very similar to the commutative case plus taking into account the
commutator of the coordinates leads to
\beq
\mathbf{F}=\gamma_{\pm}(N)\frac{i\epsilon_{abc}}{4} X_adX_b\wed dX_c,
\eeq
where
\beq
\gamma_{\pm}(N) = \frac{(1-\frac{1}{N^2})^{3/2}(N \pm (N^2 - 2))}{N^2 - 3}.
\eeq
Now for $f\in {\mathcal A}_N$ and
\[
\omega:=\frac{\epsilon_{abc}X_adX_b\wed dX_c}{8\pi} \in \Omega^2({\mathcal A}_N)
\]
define the integral trough
\beq
\int^{\star} f\omega = \tr_{\!N} f,\:\:\mbox{ where } 
\tr_{\!N}(\cdot)=\frac{1}{N}\tr(\cdot),
\eeq
with $\int^{\star}\omega =1$. The 2-form $\omega$ is our noncommutative volume 
form, in the commutative limit it converges to the normalized volume form on 
$S^2$. Consequently the first Chern number of the fuzzy line bundle $\fb$ is 
given by
\[
c_1(\fp) = \frac{1}{2\pi i} \int^{\star} \mathbf{F},
\]
and one finds
\beq
c_1(\fp) = \gamma_{\pm}(N)
\eeq
for the topological charge. In the commutative limit $\nc\rightarrow 0$,
i.e. $N\rightarrow \infty$ the topological charge is $\pm 1$.

Bundles of higher topological charge will be treated using a more general
approach in a forthcoming paper.
\section{Acknowledgements}
HG wants to thank the Center for Advanced Studies (ZHS) and the Institute for 
Theoretical Physics, Leipzig University, for the Leibniz--Professorship 
2000/2001. Both authors want to thank A.~Strohmaier for fruitful discussions 
and CWR thanks S.~Kolb for helpful remarks. This work was partly supported by 
the ``Deutsche Forschungsgemeinschaft'' within the scope of the 
``Graduiertenkolleg Quantenfeldtheorie'' of the University of Leipzig. 

\end{document}